\documentclass[twocolumn,showpacs,preprintnumbers,amsmath,amssymb,prb]{revtex4}

\usepackage{graphicx}% Include figure files
\usepackage{dcolumn}% Align table columns on decimal point
\usepackage{bm}% bold math

\begin{document}
\title{Raman modes of the deformed single-wall carbon nanotubes}

\author{Gang Wu}
\email{wugaxp@gmail.com}
\author{Jian Zhou}
\author{Jinming Dong}
\email[Corresponding author E-mail: ]{jdong@nju.edu.cn}
\affiliation{Group of Computational Condensed Matter Physics,
National Laboratory of Solid State Microstructures and Department
of Physics, Nanjing University, Nanjing 210093, P. R. China}

\begin{abstract}
With the empirical bond polarizability model, the nonresonant
Raman spectra of the chiral and achiral single-wall carbon
nanotubes (SWCNTs) under uniaxial and torsional strains have been
systematically studied by \textit{ab initio} method. It is found
that both the frequencies and the intensities of the low-frequency
Raman active modes almost do not change in the deformed nanotubes,
while their high-frequency part shifts obviously. Especially, the
high-frequency part shifts linearly with the uniaxial tensile
strain, and two kinds of different shift slopes are found for any
kind of SWCNTs. More interestingly, new Raman peaks are found in
the nonresonant Raman spectra under torsional strain, which are
explained by a) the symmetry breaking and b) the effect of bond
rotation and the anisotropy of the polarizability induced by bond
stretching.

\end{abstract}


\date{\today}
\maketitle

\section{Introduction}

Single-wall carbon nanotube (SWCNT) is a kind of nanoscale
molecule obtained by wrapping a graphene sheet into a seamless
cylinder. Since its discovery \cite{r1} in 1991 by Iijima, the
SWCNT has been of great interest in mesoscopic physics and
nanotechnology because of its exceptional electronic and
mechanical properties \cite{r2}.

At the same time, much attention has been devoted to the
vibrational properties of the SWCNTs
\cite{r3,r4,r5,r6,r7,r8,r9,r10,r11,r12,r13,r14,r15,r16}. The
excitations from the electronic density-of-state (DOS) peaks below
the Fermi level to those above it are enhanced greatly due to the
existence of the Van Hove singularities (VHS) \cite{r17} caused by
the special one-dimensional (1D) structure of the SWCNTs. As a
result, the Raman spectroscopy has been considered as a powerful
tool to detect the diameter-selective phonon modes of SWCNTs
\cite{r4,r18,r19,r20}. The low-frequency peaks in the Raman
spectrum are attributed to the radial breathing modes (RBM, $R$
band), where all carbon atoms are subject to an in-phase radial
displacement. It was found that the RBM frequency depends only on
the tube diameter in an inverse proportion to it, but not on its
chirality. On the other hand, the higher frequency peaks in the
Raman spectrum are caused by the out-of-phase vibrations of the
neighboring carbon atoms parallel to the tube surface (tangential
modes, $T$ band), which are related to the $E_{2g}$(2) phonon mode
at $\sim $1580 cm$^{ - 1}$ in graphite. In the achiral tubes, one
can distinguish two tangential modes with atomic displacements
parallel and perpendicular to the tube axis, respectively.

Beyond traditional force-constant method \cite{r8}, the
state-of-the-art first-principle methods have already been used to
calculate more accurately phonon dispersion curves
\cite{r21,r22,r23,r24,r25}, which offers a unique technique to
investigate the vibrational properties of the nanotubes or
nanoropes. Moreover, some of the first-principle methods can be
extended to predict nonresonant Raman intensities \cite{r26},
which, however, needs very intensive calculations and is less
accurate than the purely vibrational calculations, leading to
proposal of a simple empirical bond polarizability model to
calculate the Raman intensities of the fullerenes \cite{r27, r28}
and nanotubes \cite{r8}. It is proved that the method can provide
reasonable nonresonant Raman spectra in the most cases, including
the nonresonant Raman spectra of the finite-length SWCNTs
\cite{r29}.

Recently, the mechanical deformation effect on the SWCNT
electronic properties has been successfully studied by using a
simple picture based on the $\pi $ electron tight-binding
approximation \cite{r30}. Some interesting behaviors are found,
e.g., a symmetry breaking due to the deformation may remove the
degeneracy of energy levels. Experimentally, resistance of SWCNT
transistors was found to vary significantly under bending and
stretching \cite{r31}, and an observed torsion of a metallic SWCNT
was speculated to open a small band gap of the tube \cite{r32}.
Several theoretical studies on the uniaxially and torsionally
deformed SWCNTs \cite{r30, r33,r34,r35,r36} have shown that their
band structures are determined by their chiral symmetries and the
kinds of strains. Most of the deformed SWCNTs show a
metal-semiconductor transition, occurring repeatedly with
increasing strain. It is known that under uniaxial strain, the
derivative of band gap over strain is the largest for zigzag tubes
and decreases with increase of the chiral angle. In contrast,
under torsional strain, the derivative becomes the largest for
armchair tubes and decreases as the chiral angle decreases. In
particular, armchair tubes under uniaxial strain and metallic
zigzag tubes under torsional strain remain to be metallic
\cite{r30}.

At the same time, the deformation effect on the SWCNT vibration
properties, especially its Raman spectroscopy, has attracted much
attention in experiments \cite{r37,r38,r39,r40,r41,r42,r43} and
theoretical studies \cite{r12, r45}. Kahn \textit{et al.}
\cite{r12} studied the hydrostatic pressure dependence of the
Raman modes for different nanotubes, and found that their shifts
under pressure differ from the corresponding ones of graphite. It
is found in Refs. \onlinecite{r37, r42, r43} that the RBM shifts
towards higher wavenumbers with increasing hydrostatic pressure at
a rate of 7$\sim $10 cm$^{ - 1}$GPa$^{ - 1}$. With a generalized
tight-binding molecular dynamics (GTBMD) scheme, Venkateswaran
\textit{et al.} \cite{r42} analyzed the effect of the intertube
Van der Waals interaction on the $R$ band shift. And Refs.
\onlinecite{r37} and \onlinecite{r38} reported that before a
structural transition at near 2 GPa, the tangential modes
exhibited essentially the same pressure dependence of 5.7-5.8 and
5.3-6.1 cm$^{ - 1}$GPa$^{ - 1}$, respectively. More details about
this hydrostatic pressure effect on the Raman modes can be found
in the review article \cite{r45}. Nevertheless, all of these
researches focuses on the hydrostatic pressure effect, and the
uniaxial and torsional strain effects have not been studied.

In this work we will study systematically the nonresonant Raman
spectra of chiral and achiral SWCNTs under uniaxial and torsional
strains. The Raman intensity is calculated by using the
eigenvectors of the vibrational modes obtained by \textit{ab
initio} method, and the polarizability parameters are obtained by
using the empirical bond polarizability model \cite{r28, r8}. The
detail of methodology is presented in Sec. II. The numerical
results are discussed in Sec. III. Finally, this work ends with
some conclusions in Sec. IV.

\section{MODEL AND METHOD}

\subsection{Cumulant force-constant method}

In most of published papers on the phonon dispersions of the
SWCNTs obtained by \textit{ab initio }calculations, the
low-frequency dispersions were not well reproduced mainly due to
use of the interatomic force-constant (IFC) method. In Ref.
\onlinecite{r25}, by means of Cumulant Force Constant (CFC)
method, L. H. Ye \textit{et al.} successfully calculated phonon
dispersions of the nanotubes and graphite. Details about
comparison between the CFC and IFC methods are available in their
original paper \cite{r25}. In this work, we use CFC method to
construct the dynamical matrices for both of the undeformed and
deformed SWCNTs, and then calculate their phonon dispersion
curves.

First, we summarize the reason of employing CFC method in this
work:

a). The 1D-like geometrical structure of the carbon nanotube makes
its physical properties highly anisotropic, for which it is
definitely necessary to use the CFC instead of the IFC.
Comparatively, the CFC is also an approximated method, but it
keeps the most rotational symmetries, which means although the
numerical values of the dynamical matrices are obtained
approximately, the symmetries of the dynamical matrices are
correct. This is the reason why it works so well for an
anisotropic system.

b). In the IFC method, in order to minimize the truncation effect,
one should use a large supercell, spending much computational
time. At the same time, the numerical noise in force calculations
may become very serious if too large supercell is used, which may
reduce the accuracy of phonon calculation. On the other hand,
because the CFC method does not need to perform an explicit
truncation on the interatomic interactions, as shown in Ref.
\onlinecite{r25}, so, only a very small supercell (1x1x2) in the
CFC method is sufficient to calculate accurately the phonon modes
on the $\Gamma$ point and the $X$ point at Brillouin zone
boundary. This also means that the phonon modes of even achiral
SWCNTs, such as (12, 4) tube containing 208 atoms in its one unit
cell, can be efficiently calculated under current hardware
conditions.

\subsection{Empirical bond polarizability model}

The time-averaged power flux of the Raman-scattered light at a
given direction of a solid angle $d\Omega $ in a frequency range
between $\omega _f $ and $\omega _f + d\omega _f $ is related to
the differential scattering cross section as follows,

\begin{widetext}
\begin{equation}
\label{eq1} \frac{d^2\sigma }{d\Omega d\omega _f } =
\frac{1}{8\pi^2c^2}\omega _f^3 \omega _i \left[ {\left\langle
{n\left( \omega \right)} \right\rangle + 1} \right]\hbar \times
\sum\limits_{\alpha \beta \gamma \lambda } {v_\alpha v_\beta
H_{\alpha \gamma \beta \lambda } \left( \omega \right)w_\gamma
w_\lambda } ,
\end{equation}

\end{widetext}

\noindent where

\begin{equation}
\label{eq2}
 H_{\alpha \gamma \beta \lambda } \left( \omega \right)= \\
 \sum\limits_j {a_{\alpha \gamma }^ * \left( j
\right)a_{\beta \lambda } \left( j \right)\frac{1}{2\omega _j
}\left[ {\delta \left( {\omega - \omega _j } \right) + \delta
\left( {\omega + \omega _j } \right)} \right]} ,
\end{equation}

\noindent with

\begin{equation}
\label{eq3} a_{\alpha \gamma } \left( j \right) =
\sum\limits_{n\delta } {\frac{\pi _{\alpha \gamma ,\delta }^n
}{\sqrt {M_n } }\left\langle {\delta n} \mathrel{\left| {\vphantom
{{\delta n} j}} \right. \kern-\nulldelimiterspace} {j}
\right\rangle } .
\end{equation}

Here, $\omega _i $ and $\omega _f $ are the frequencies of the
incident and scattered light; $\omega \equiv \omega _i - \omega _f
$ is the Raman shift.
$\mathord{\buildrel{\lower3pt\hbox{$\scriptscriptstyle\rightharpoonup$}}\over
{v}} $ and
$\mathord{\buildrel{\lower3pt\hbox{$\scriptscriptstyle\rightharpoonup$}}\over
{w}} $ are the corresponding polarization unit vectors of the
incident and scattered light, respectively. $\left\langle {n\left(
\omega \right)} \right\rangle $ is the Bose factor, $M_n $ is the
mass of the $n$th atom, and $\omega _j $ and $\left\langle {\delta
n} \mathrel{\left| {\vphantom {{\delta n} j}} \right.
\kern-\nulldelimiterspace} {j} \right\rangle $ are the frequency
and $\left( {\delta n} \right)$ component of the $j$th mode. The
coefficient $\pi _{\alpha \gamma ,\delta }^n $ in Eq. (\ref{eq3})
connects the polarization changes to the atomic motions
\cite{r48}, which is obtained by expanding the polarizability
tensor $\pi ^n$ in terms of the atom displacements $u_\delta ^n $,
with

\begin{equation}
\label{eq4} \pi _{\alpha \gamma ,\delta }^n = \sum\limits_m
{\left( {\frac{\partial \pi _{\alpha \gamma }^m }{\partial
u_\delta ^n }} \right)_0 } .
\end{equation}

Adopting the nonresonant empirical bond-polarization theory
\cite{r27,r28, r8}, the Eq. (\ref{eq4}) can be rewritten as:

\begin{widetext}
\begin{equation}
\label{eq5} \pi _{\alpha \beta ,\gamma }^n = \sum\limits_m
{\frac{\left( {2{\alpha }'_\parallel + {\alpha }'_ \bot }
\right)}{3}\delta _{\alpha \beta } \hat {r}_\gamma + \left(
{{\alpha }'_\parallel - {\alpha }'_ \bot } \right)\left( {\hat
{r}_\alpha \hat {r}_\beta - \frac{1}{3}\delta _{\alpha \beta } }
\right)\hat {r}_\gamma + \frac{\left( {\alpha _\parallel - \alpha
_ \bot } \right)}{r}\left( {\delta _{\alpha \gamma } \hat
{r}_\beta + \delta _{\beta \gamma } \hat {r}_\alpha - 2\hat
{r}_\alpha \hat {r}_\beta \hat {r}_\gamma } \right)} .
\end{equation}
\end{widetext}

Here, $\hat {r}$ is the unit vector of the vector $\vec{r}$
connecting the $n$ and $m$ atoms linked by a bond.
$\alpha_{\parallel}$ and $\alpha_{\bot}$ represent the static
longitudinal and perpendicular bond polarizability, respectively,
and ${\alpha }'_\parallel = \left( {\frac{\partial \alpha
_\parallel }{\partial r}} \right)_0 $, ${\alpha }'_ \bot = \left(
{\frac{\partial \alpha _ \bot }{\partial r}} \right)_0 $. The
values of $\alpha _\parallel $, $\alpha _ \bot $, ${\alpha
}'_\parallel $ and ${\alpha }'_ \bot $ are given empirically as a
function of the bond length between two carbon atoms. For
undeformed SWCNTs, there is only single type of bonds and the bond
polarizability model is completely defined by three parameters
\cite{r8}: $\alpha = 2{\alpha }'_\parallel + {\alpha }'_ \bot =
4.7$ {\AA}$^{2}$, $\beta = {\alpha }'_\parallel - {\alpha }'_ \bot
= 4.0$ {\AA}$^{2}$, and $\gamma = \alpha _\parallel - \alpha _
\bot = 0.04$ {\AA}$^{3}$. It is known, however, that the
polarizability parameters of carbon are similar for a variety of
carbon materials. Furthermore, the relative intensities of the
Raman modes are not so sensitive to small changes of the bond
polarization parameter values, except for the lowest $E_{2g} $
mode. Thus in the present work, we still use the above
polarization parameters for the deformed SWCNTs, which can only
lead to a small error in the peak height, and does not affect the
peak positions. In fact, Guha et al. \cite{r28} chose different
bond polarizability parameters to calculate the Raman spectra of
the fullerenes, causing only small changes of the relative Raman
intensities, which is also found in our calculations. For example,
even use of a set of parameters obtained by interpolating those of
the fullerenes and nanotubes does not bring any qualitative
changes of our results.

In all our calculations, the line shape of each peak is assumed to
be Lorentzian and its line-width is fixed at 1.68 cm$^{ - 1}$. The
tube axis is taken along the $Z$ direction, and the $X$ and $Y$
directions are taken to be perpendicular to it. In Ref.
\onlinecite{r8}, two possible geometries for the light
polarization are mentioned: the \textit{VV} and \textit{VH}
configurations. In the \textit{VV} configuration, the incident and
the scattered polarizations are parallel to each other, while they
are perpendicular to each other in the \textit{VH} configuration.
After careful calculations, we can find all of the Raman-active
peaks in the \textit{VV} configuration. For the \textit{VH}
configuration, there is no any qualitative change of our results
compared with those obtained from \textit{VV} configuration. Due
to this reason, and also because we pay our main attention to the
shift of the Raman mode frequency under applied strains, only the
results in the \textit{VV} configuration have been presented in
this work.

\subsection{Calculation details}

Our first principle calculations are based on the total energy
plane-wave pseudopotential method in the local density
approximation (LDA). The Vienna \textit{ab initio} simulation
package (VASP) \cite{r49, r50} is employed in this paper. In the
most recent version of VASP, the interaction between the ions and
electrons is described by the projector augmented-wave \cite{r51}
(PAW) in the implementation of Kress and Joubert \cite{r52}. The
PAW method is always considered to be able to give a more accurate
and reliable result than the ultrasoft potential. In the present
PAW potential, the 2$s$ and 2$p$ orbitals are treated as valence
orbitals, and a plane-wave cutoff of 400 eV is used to obtain
reliable results.

A supercell geometry was adopted so that the nanotubes are aligned
in a hexagonal array with the closest distance between adjacent
nanotube surfaces being at least 10 \AA, which is found to be
large enough to prevent the tube-tube interactions.

The K-points sampling in the reciprocal space is a uniform grid
$(1{\times}1{\times}n)$ along the nanotube axis ($Z$ direction) in
our calculations, the maximum spacing between $k$ points is 0.03
\AA$^{-1}$ and the Gaussian smearing width is 0.03 eV. This means
that $n$ is obtained by dividing 30 with the supercell length
along the tube axis. For example, for a supercell
$(1{\times}1{\times}2)$ constructed by the unit cell of a zigzag
tube, the $n$ equals to $30/(T_{zig}*2)\thickapprox6$, where
$T_{zig}\thickapprox2.451$ \AA$ $ is the unit cell length of the
zigzag tube. The increasing K-points sampling has no significant
effect on the frequencies of the final Raman active modes, e.g.,
doubling the number of special k points only causes less than 4
cm$^{-1}$ change on the most $\Gamma$ point phonon frequencies.

All SWCNT geometrical structures used in this work have been fully
optimized before performing further force constants (FCs)
calculation. The structural optimization process is almost the
same as Ref. \onlinecite{r25}. Finally, the optimal structure is
obtained when the residue forces acting on all the atoms were less
than 0.005 eV/\AA$ $ for achiral SWCNTs and 0.01 eV/\AA$ $ for
chiral SWCNTs. The equilibrium lattice constants and averaged
nanotube radii are listed in Table I. Our structure parameters of
(5, 5), (10, 0) and (10, 5) tubes are almost the same as those in
Ref. \onlinecite{nr44}, which are also obtained by VASP.

Due to the high symmetry of an ideal SWCNT, so there exists only
one inequivalent atom. Thus the FCs of one atom are sufficient to
construct the force constant matrix of the whole tube by applying
the tube symmetry.

\begin{table}[htbp]
\caption{\label{table1} Optimized structural parameters of the undeformed
nanotubes. $R_0 $ is the average radius, and $T_0 $ is the lattice
constant along the tube axis.}
\begin{ruledtabular}
\begin{tabular}{ccc}
Tubes& $R_0 ( {\AA} )$& $T_0
( {\AA} )$ \\
\hline (6, 0)& 2.397&
4.230 \\
(7, 0)& 2.773&
4.235 \\
(8, 0)& 3.159&
4.233 \\
(9, 0)& 3.541 &
4.237 \\
(10, 0)& 3.927&
4.237 \\
(11, 0)& 4.313 &
4.238 \\
(5, 5)& 3.408 &
2.448 \\
(6, 6)& 4.075&
2.448 \\
(10, 5)& 5.173&
11.21 \\
(12, 4)& 5.636&
15.28 \\
\end{tabular}
\end{ruledtabular}
\end{table}

\section{Numerical results and discussions.}

In this section, the Raman spectra of the infinite-length deformed
SWCNTs are investigated. Firstly, the tensile strain effect will
be discussed in Sec. III A, and the torsional strain one will be
given in Sec. III B.

\subsection{Tensile strain}

The phonon dispersion relations and density of states for zigzag
(10, 0) tube are shown in Fig. 1, from which, one can find that
the low-frequency dispersions are well produced, and when the
tensile strain is applied, the low-frequency vibration modes
almost do not change, while the high-frequency vibrations show
sensitive response to the tensile strain.

\begin{figure*}
\includegraphics[width=0.66\columnwidth]{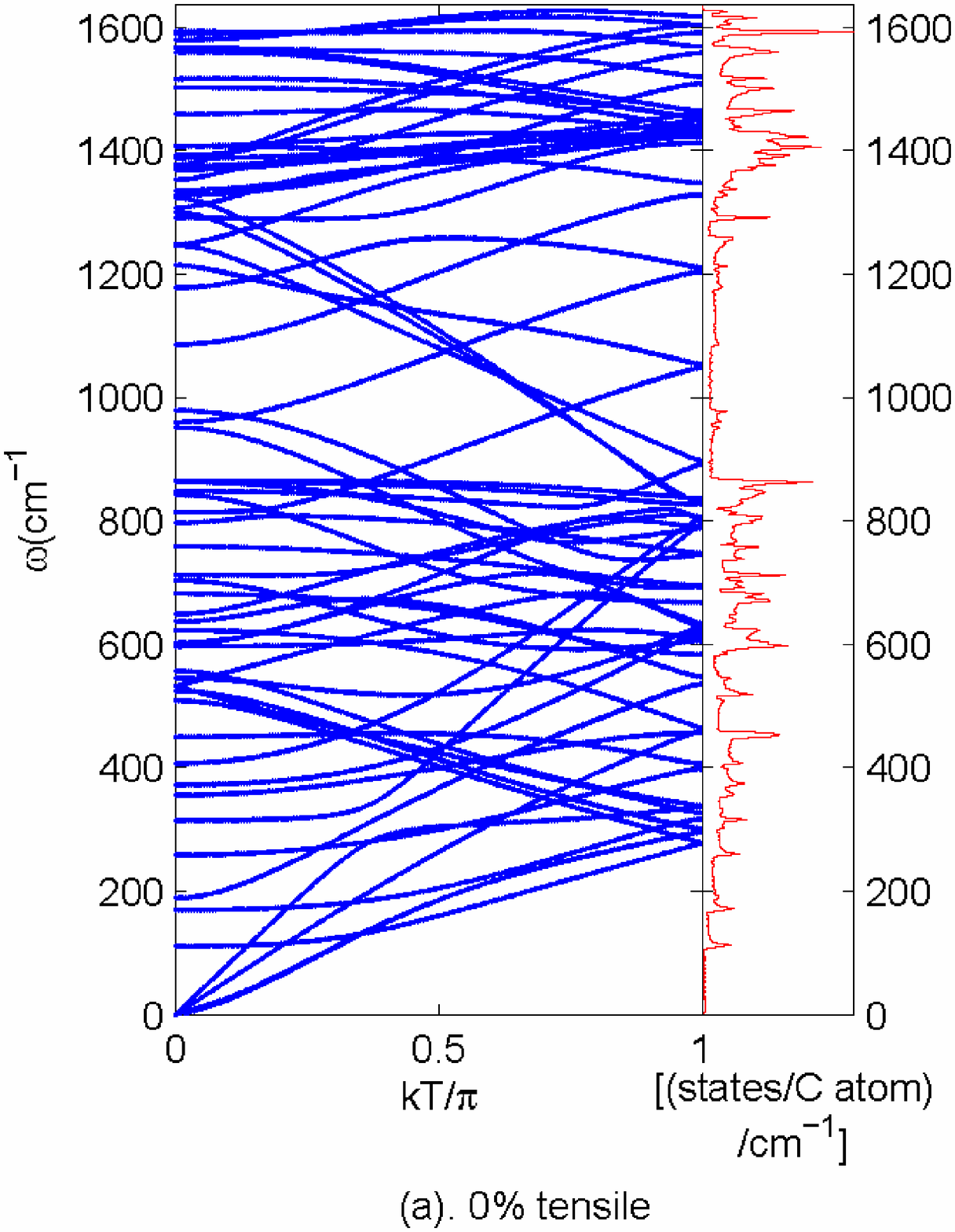}
\includegraphics[width=0.66\columnwidth]{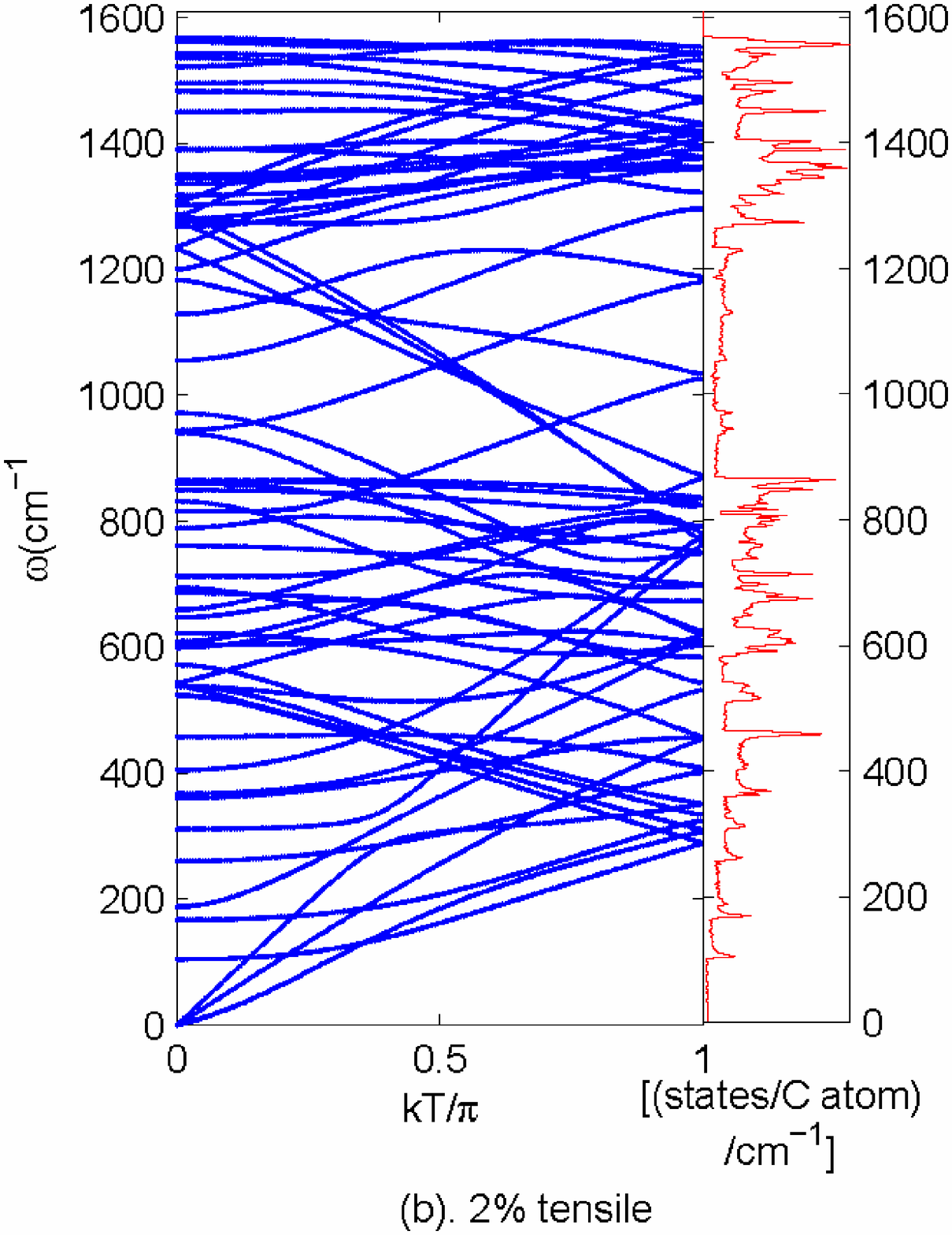}
\includegraphics[width=0.66\columnwidth]{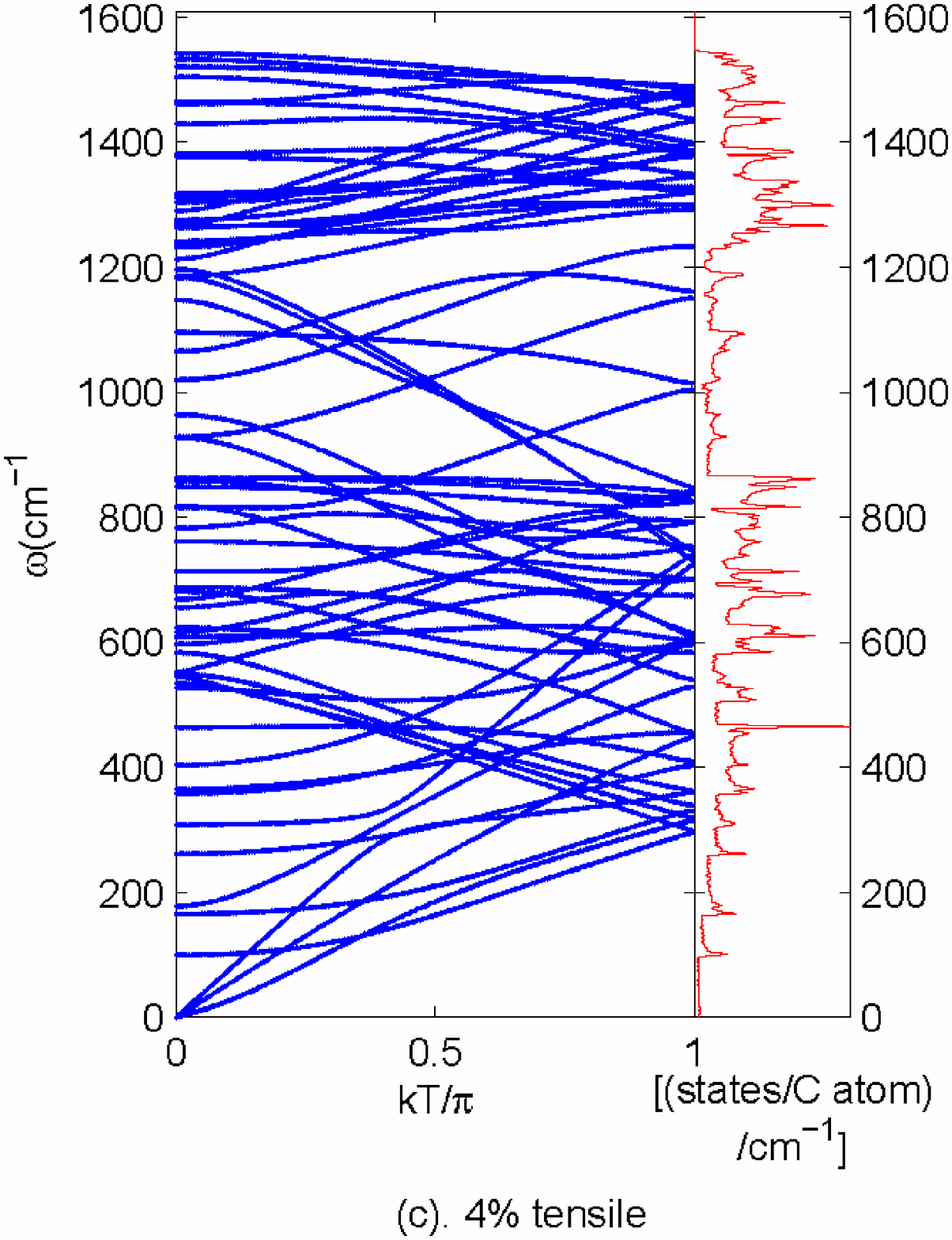}
\caption{\label{fig1}Calculated phonon dispersion curves and
density of states are shown in the left and right panels,
respectively, in each sub-figure. For (10, 0) zigzag nanotube
under the tensile strains of 0{\%}, 2{\%} and 4{\%}.}
\end{figure*}

These characters can be seen more clearly in the Raman spectra.
First we will consider the Raman spectra of zigzag tubes as an
example. Because the cylindric structure and the neighboring
relationships do not change under the tensile strain, we find the
intensities of the Raman active modes almost do not change at all.
Thus only the frequency shifts for the Raman active modes under
tensile strain are given in Fig. 2. In the present work, the
tensile strain is defined as $e = (T / T_0 ) - 1$, where $T$ and
$T_{0}$ are the lengths of the axial unit cells of the deformed
and undeformed nanotubes, respectively. The values of $T_{0}$ have
been given in Table I. After applying tensile strain, only the
positions of the carbon atoms are relaxed to obtain a reliable
optimized structure. It can be seen clearly from Fig. 2 that when
the tensile strength increases: a) The frequencies of the
low-frequency Raman active modes almost do not change. While the
high-frequency part of the Raman spectra is shifted downward,
which consists with the available experimental and theoretical
results \cite{r37,r38,r39,r40,r41,r42,r43,r45, r12}. b) The
deformation does not change the number of the Raman active
vibrations. c) The shifting rate of the Raman modes is almost the
same for the nanotubes with different radii, i.e., they shift
linearly.

\begin{figure*}
\includegraphics[width=0.66\columnwidth]{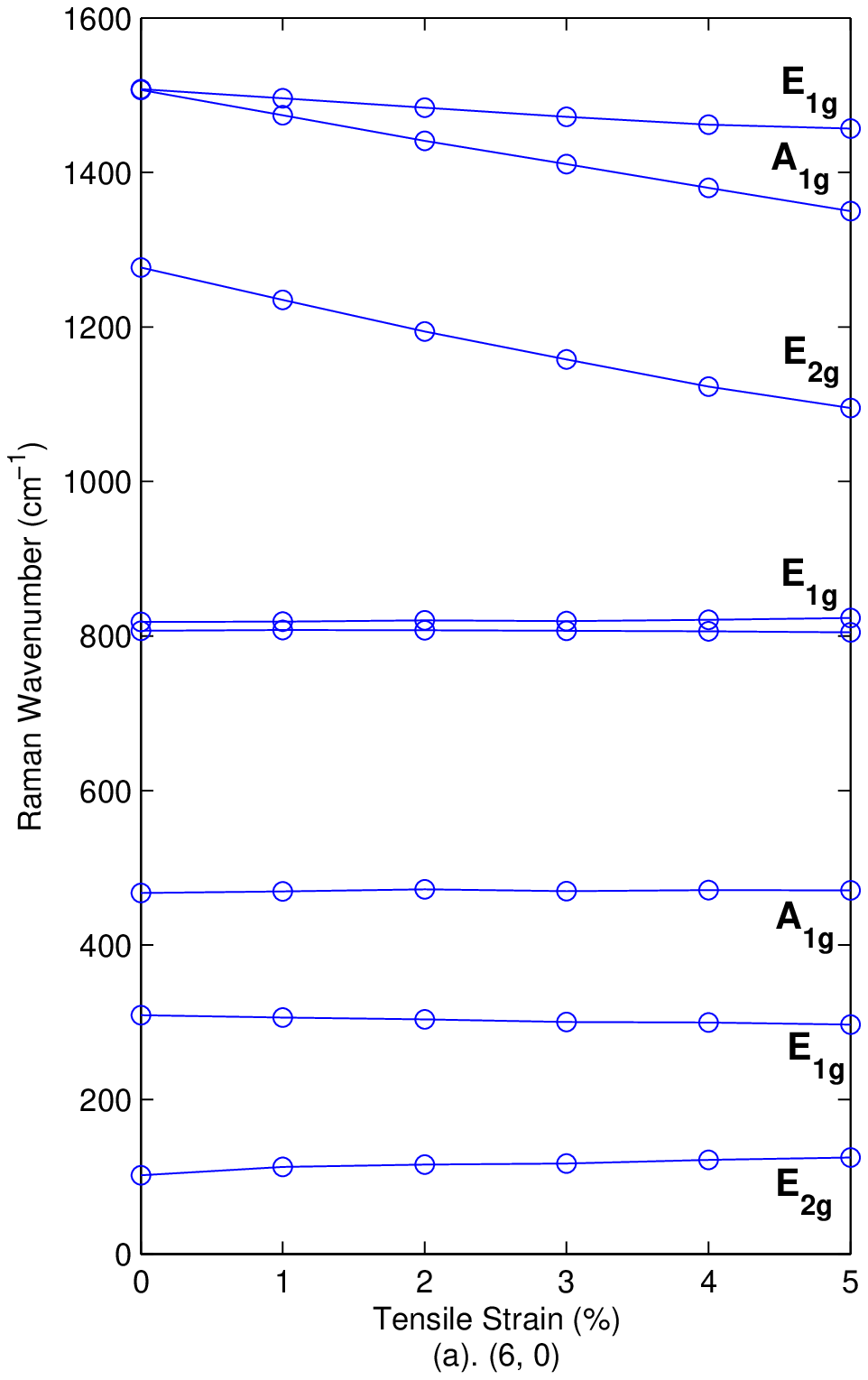}
\includegraphics[width=0.66\columnwidth]{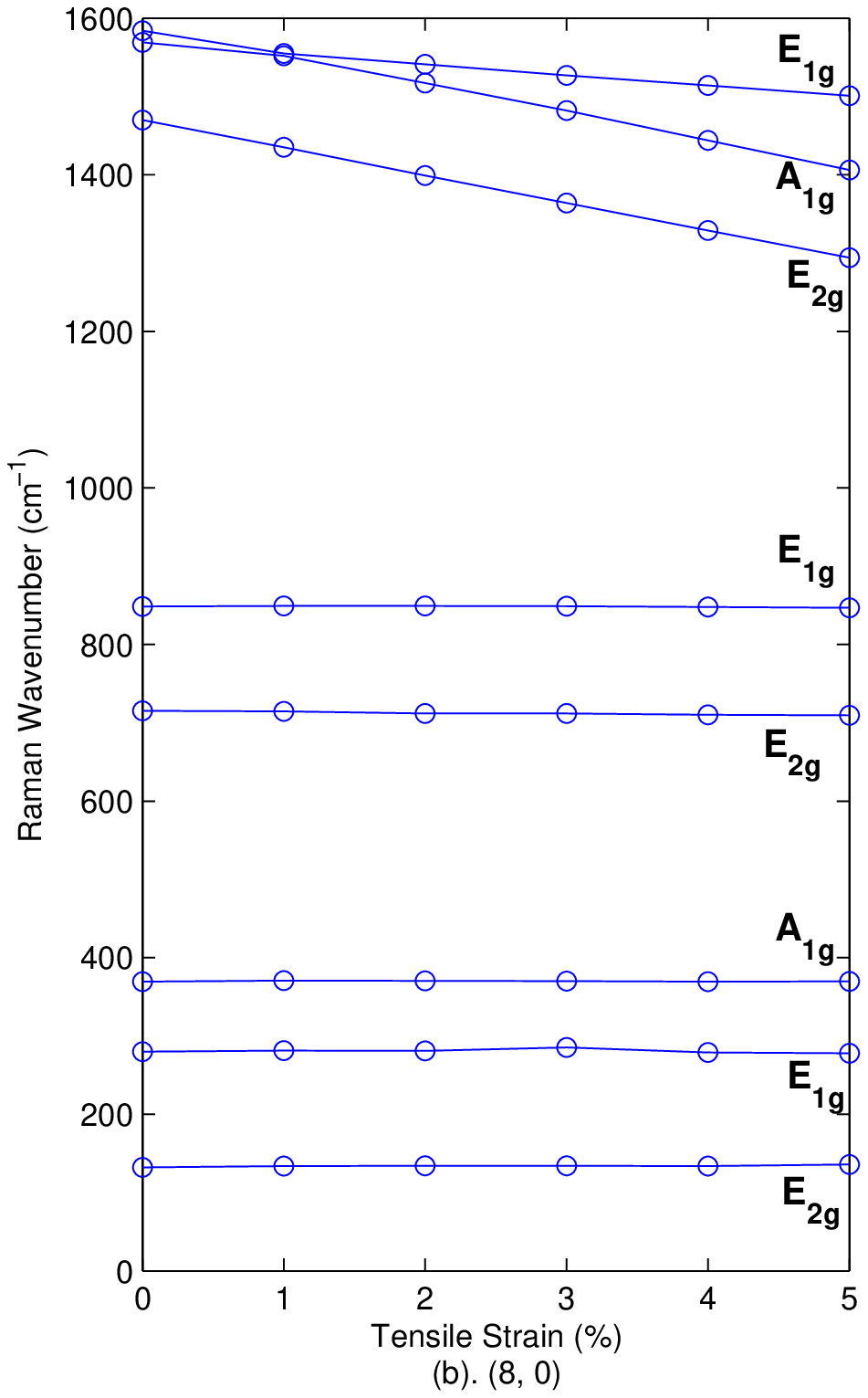}
\includegraphics[width=0.66\columnwidth]{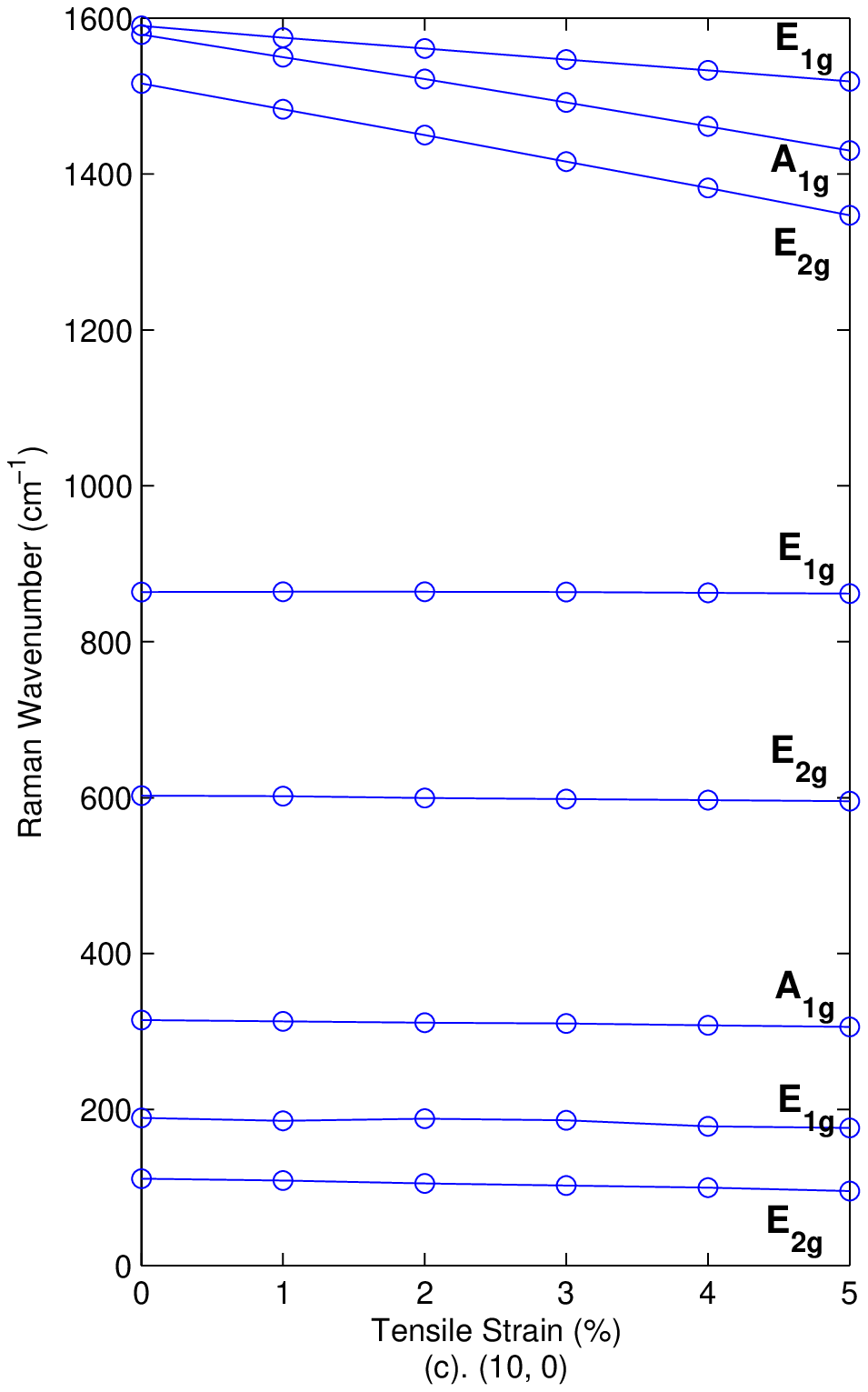}
\caption{\label{fig2}Frequencies of Raman active modes for (n, 0)
zigzag tubes (n=6, 8 and 10) under tensile strain.}
\end{figure*}

In fact, above three results are reasonable and can be understood
as follows. Because different nanotubes have almost the same
Poisson ratio $\sigma $ of 0.2, the tube radius only changes about
1{\%}, even if the tube is elongated by 5{\%}. And for lower Raman
modes, the vibrational frequencies are mainly affected by the tube
diameter, making their peak positions do not move much.
Especially, for the radial breathing mode (RBM), which is found to
be inversely proportional to the tube diameter, only very small
shift (about 1 cm$^{ - 1})$ is found when the tensile strength
increases up to 5{\%}.

But the bonds parallel to the tube axis will be elongated when
tensile strain is applied. For higher Raman frequencies, the
vibrations are sensitive to the local bond structure, especially
its length. When bond length is elongated, the interaction among
nearby atoms will become weaker, making the Raman frequency
smaller.

Now we would like to discuss why the number of Raman active modes
does not change under the tensile strain. Following the discussion
of Ref. \onlinecite{r53}, the number of Raman active modes can be
determined by using the nonsymmorphic rod-groups of SWCNTs. All
achiral carbon nanotubes can be shown by this method to possess
only 8 Raman-active phonon modes \cite{r53}, and for all chiral
carbon nanotubes, 14 Raman-active modes can be found. Furthermore,
for the first-order Raman scattering, the point group is
sufficient because this process does not change the wave vector
$k$. Another fact is that the point groups of achiral (zigzag or
armchair) and chiral SWCNTs are $D_{2nh}$ and $D_{N}$,
respectively, where $n$ is the tube index and $N$ is the number of
graphene hexagons in the nanotube unit cell. These point groups do
not change under tensile strain. Correspondingly, simply following
the discussion in Ref. \onlinecite{r53}, one can conclude that
when tensile strength increases, the Raman active vibration number
of zigzag nanotubes will keep 8, which has been shown by our
numerical results.

Then let us analyze the detail of frequency shift of Raman active
modes. From Fig. 2, one can find that the modes below 1000 cm$^{ -
1}$ is insensitive to the tensile strain. And above 1000 cm$^{ -
1}$, the frequencies of Raman active modes decrease linearly with
tensile strain, for which there exist two different decreasing
slopes for: (i). E$_{1g}$, and (ii). E$_{2g}$ and A$_{1g}$ modes.

\begin{figure}
\includegraphics[width=0.8\columnwidth]{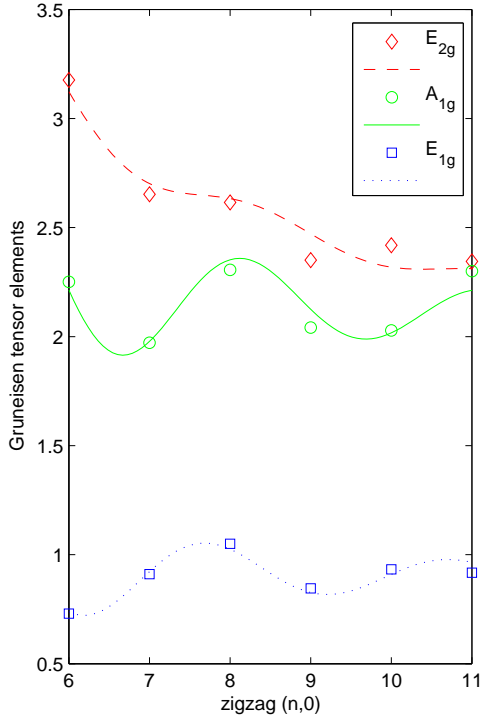}
\caption{\label{fig3}(Color online) The GTEs for the highest three
Raman active modes of zigzag tubes $(n, 0)$, $(n=6,\ldots,11)$.
The lines are fitting results (see text), and other markers
(circles, diamonds and squares) are GTEs obtained by
first-principle calculations.}
\end{figure}

Similar phenomenon has been found in the graphite system
\cite{r54}. Following Ref. \onlinecite{r54}, we also calculated
the Gr\"{u}neisen tensor elements (GTEs) $\kappa = - \frac{d\ln
\omega }{d\ln (1 + e)}$, where $\omega $ is the frequency of Raman
active modes, and $e$ is the tensile strain. We plot the GTEs vs
the tube index $n$ of the zigzag tubs in Fig. 3, and further fit
them with a same fitting function in the same plot. The fitting
function is obtained based on the following factors: a) The GTEs
must have the same dependence on the tube radii as the frequencies
of Raman modes. In Ref. \onlinecite{nr55}, the higher Raman
frequencies depend on the tube radius via the confined wave vector
$q$ around the equatorial direction, where
$q\varpropto{1/r}\varpropto{1/n}$. Here we following this idea and
set the nonoscillating part of the fitting function as
$A+\frac{B}{n^2}$, where $A$ and $B$ are undetermined
coefficients, whose quadratic form fulfils the phonon dispersion
near the $\Gamma$ point of graphite. b) The above $\kappa $ form
still can not describe the oscillating behavior of the GTEs. Thus
an extra oscillating term must be added into the fitting function,
which can be written as
$\frac{C}{n^2}{\sin(\frac{2\pi}{3}n+\phi_0)}$, where the period
has been set as 3 to match the oscillating period of the energy
gap of zigzag tubes, and the denominator $n^2$ is used to let this
term quickly vanish when $n\rightarrow\infty$. So the final
fitting function can be written as: $\kappa=A+ \frac{B}{n^2}+
\frac{C}{n^2}{\sin(\frac{2\pi}{3}n+\phi_0)}$. After making a
simple least-squares fit, finally we can obtain $\kappa _{E_{1g} }
\left( {r \to \infty } \right) = \mbox{0.897}$, $\kappa _{A_{1g} }
\left( {r \to \infty } \right) = \mbox{2.05}$, and $\kappa
_{E_{2g} } \left( {r \to \infty } \right) = \mbox{1.95}$, which
can be checked by future experiments. This also offers a powerful
tool to determine the strength of the tensile strain, i.e., after
detecting the Raman-active modes under undeformed and deformed
conditions, one can easily estimate how big the applied tensile
strain is.

Next, let us consider the armchair and chiral SWCNTs. We find
those three key characters found for the Raman spectra of zigzag
tubes are also hold for these SWCNTs. The reason should be the
same as that for the zigzag tubes. Because we mainly concern the
frequency shift, so the Raman spectra will not be shown, and
instead only the frequencies of the Raman active modes for some
armchair and chiral tubes under tensile strains are shown in Fig.
4, which can already represent all the key characters clearly.

\begin{figure*}
\includegraphics[width=0.66\columnwidth]{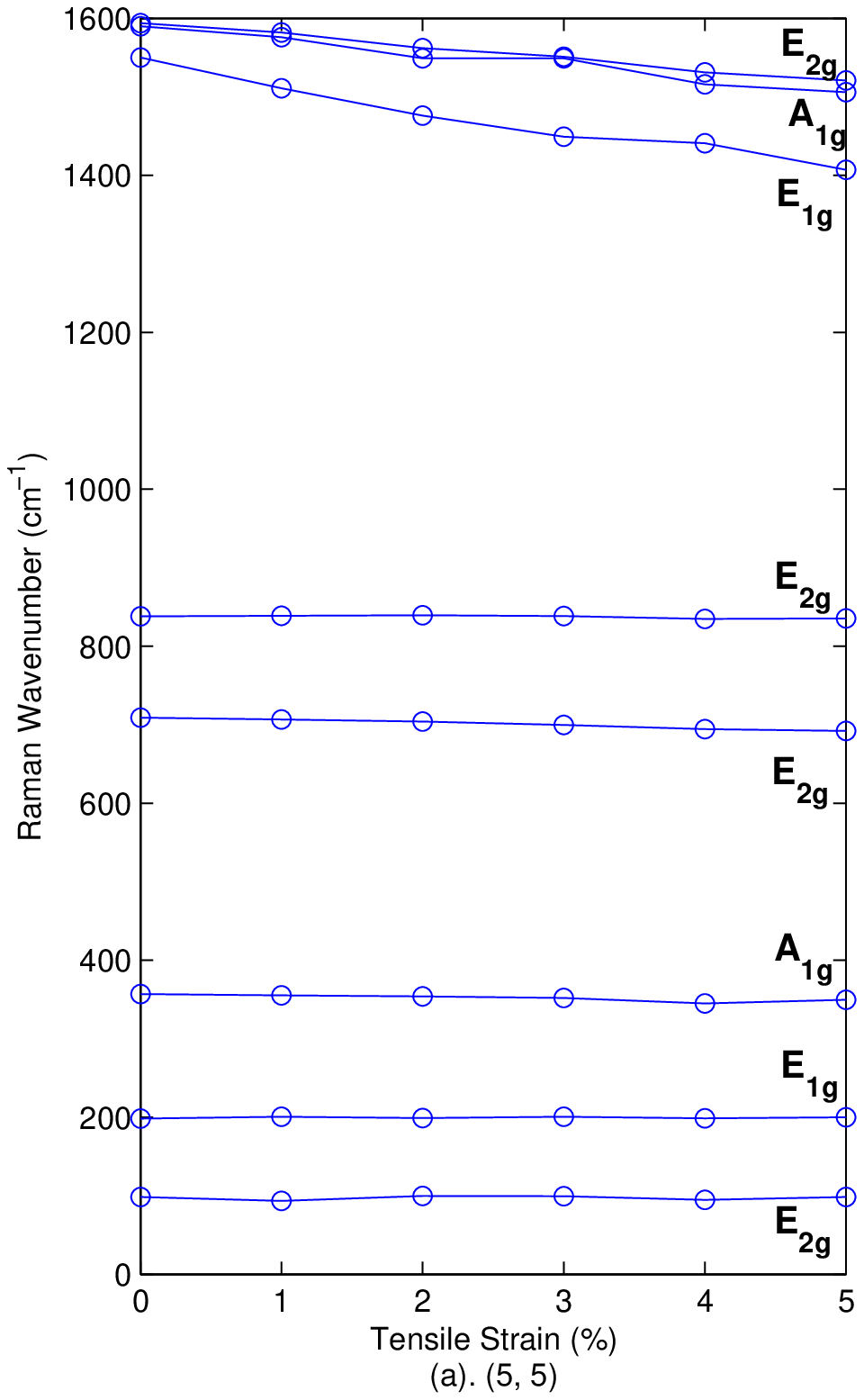}
\includegraphics[width=0.66\columnwidth]{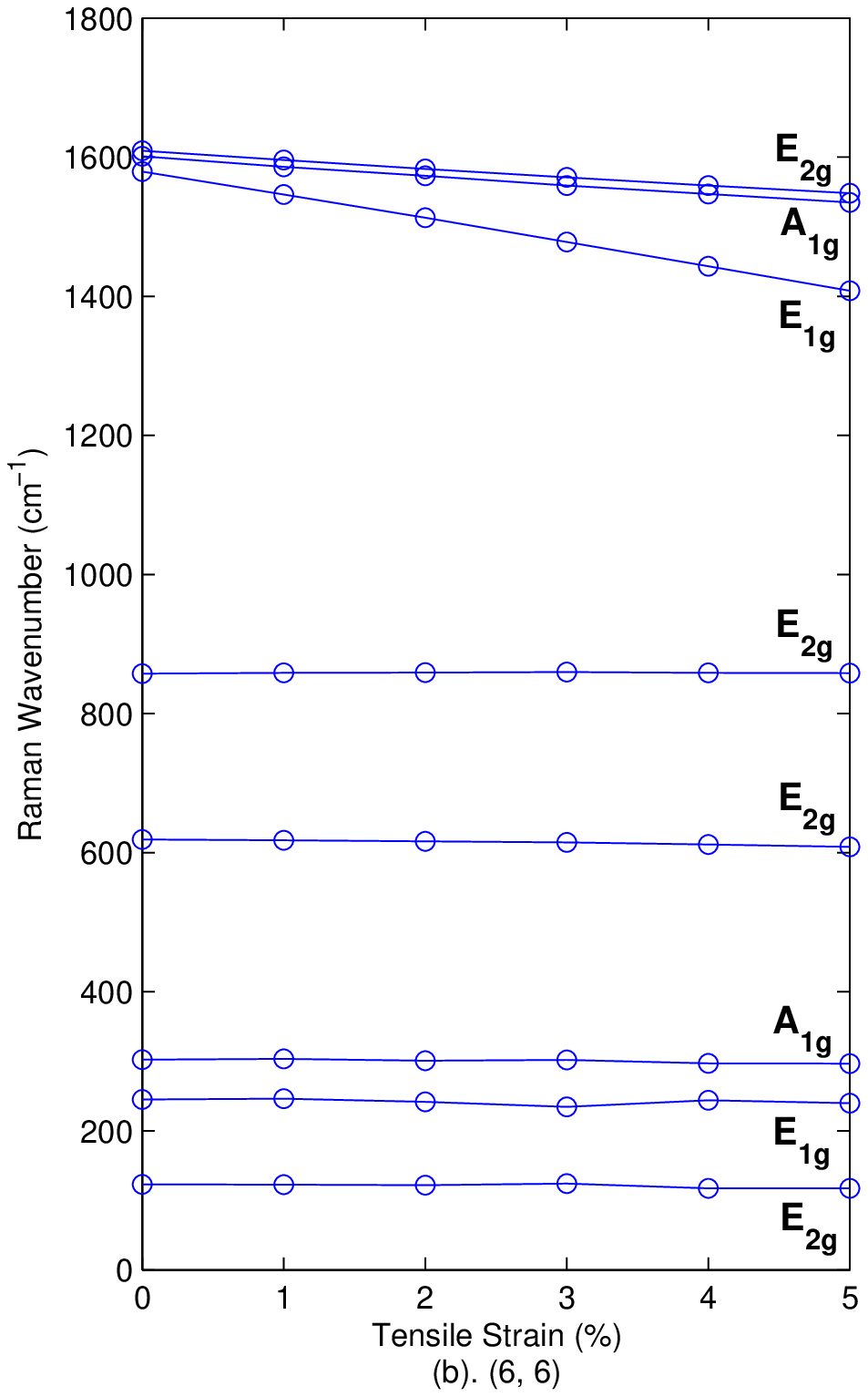}
\includegraphics[width=0.66\columnwidth]{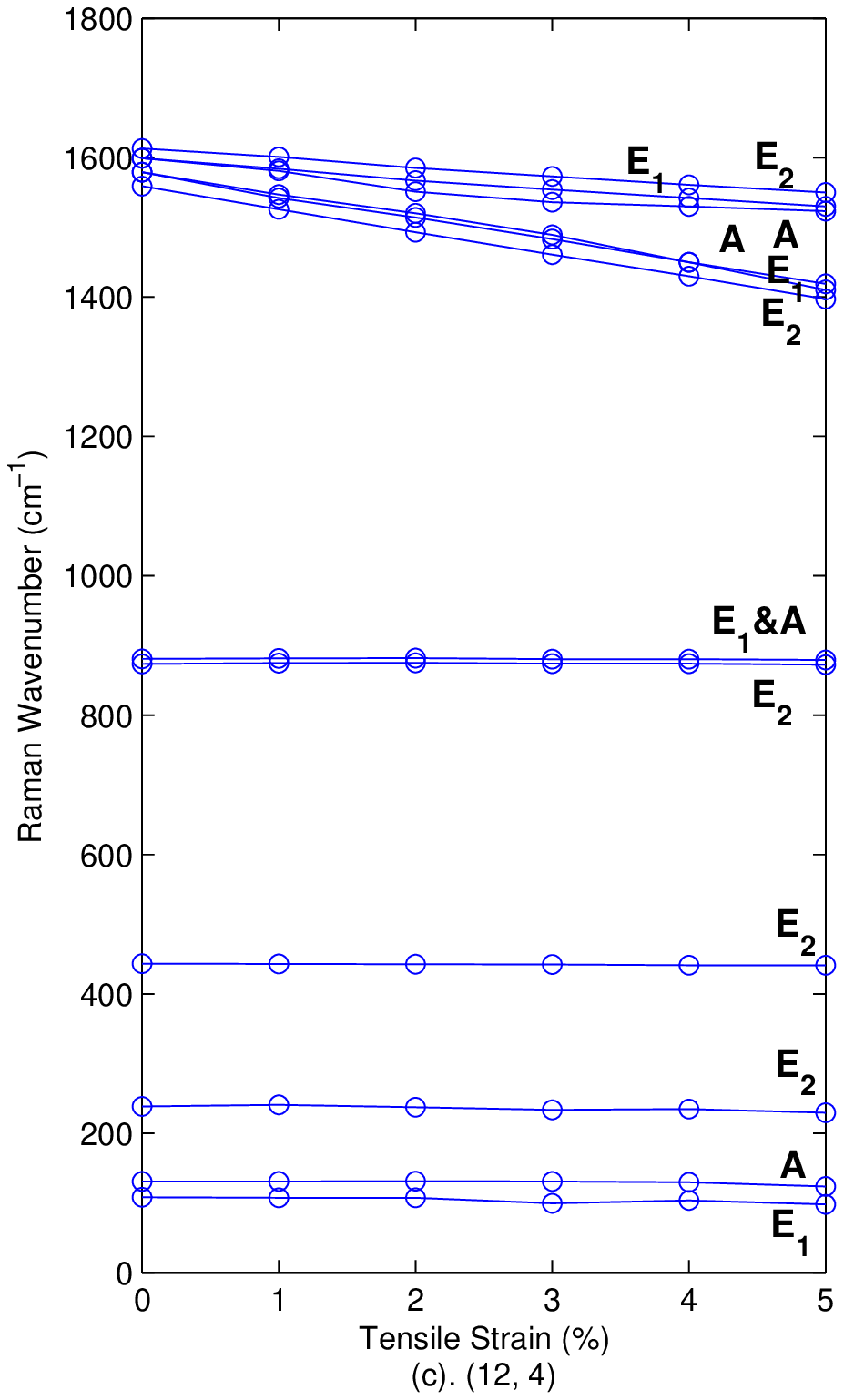}
\caption{\label{fig4}Frequencies of Raman active modes for some
armchair and chiral SWCNTs under tensile strain. (a) (5, 5), (b)
(6, 6), (c) (12, 4).}
\end{figure*}

The GTEs for these tubes are also calculated. It is interesting to
note that the GTEs of the high-frequency Raman modes of armchair
tubes can also be divided into two groups, and their values are
very close to those of the zigzag tubes. For example, for (6, 6)
tube, $\kappa _{E_{1g} } = \mbox{2.35}$, $\kappa _{A_{1g} } =
\mbox{0.861}$, and $\kappa _{E_{2g} } = 0.794$. And for (12, 4)
tube, the GTEs of the six highest frequency Raman modes are: 2.25,
2.27, 2.28, 0.994, 0.907 and 0.846. So, we suppose that for any
SWCNTs, there are only two kinds of GTEs for the high-frequency
Raman modes. In fact, detailed analysis shows that those modes
with the larger GTEs are the tangential with atomic displacements
parallel to the tube axis, and those modes with the smaller GTEs
are the tangential modes with atomic displacements perpendicular
to the tube axis. This fact could be clearly seen from Fig. 5,
showing the atomic displacements of three high-frequency
Raman-active tangential modes of (6, 6) tube.

\begin{figure}
\includegraphics[width=\columnwidth]{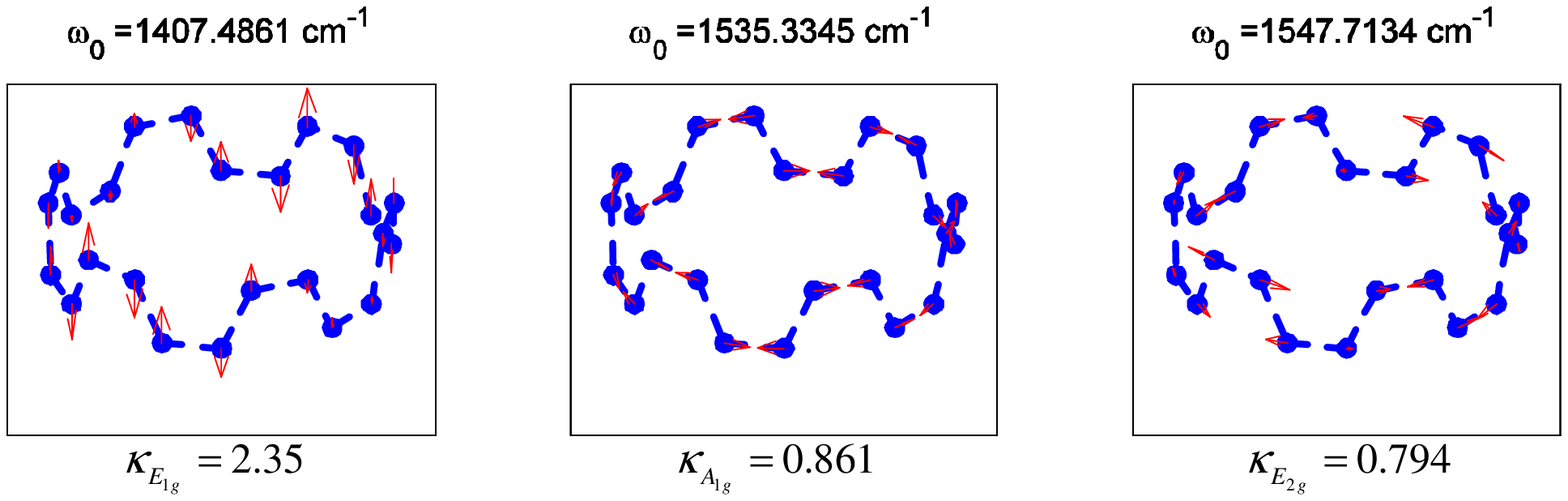}
\caption{\label{fig5}(Color online) The atomic displacements of
three Raman-active tangential modes of (6, 6) tube. $\omega _0 $
is the frequency of Raman-active modes of undeformed tube.}
\end{figure}

Finally, one may notice that it is different for the tensile
strain to affect the electronic and vibrational properties of
SWCNTs. As mentioned in the introduction part, most of the
deformed SWCNTs show a metal-semiconductor transition, occurring
repeatedly with increasing strain. Correspondingly, the linear
optical responses of the deformed carbon nanotubes also show
periodical change with increasing strain \cite{r55}. But as shown
in our calculations, the frequencies of Raman active modes change
monotonously with applied tensile strain and their intensities
only change very little, which is consistent with the available
experiments \cite{r37,r38,r39,r40,r41,r42,r43,r56}.

\subsection{Torsional strain}

\begin{figure}
\includegraphics[width=0.8\columnwidth]{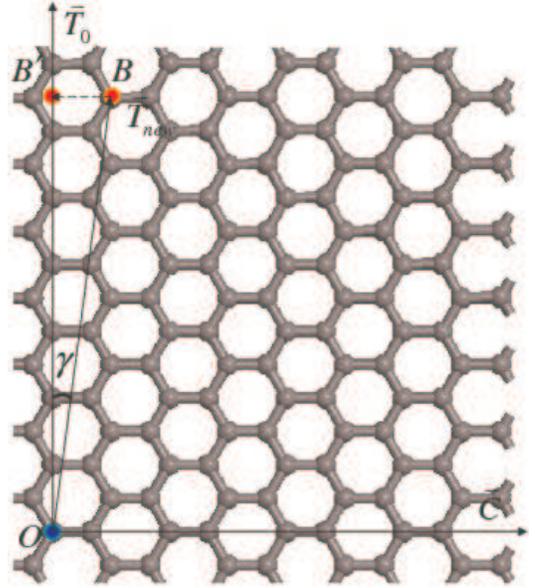}
\caption{\label{fig6}(Color online) 2D projection of the nanotubes
under the torsional strain, where $\gamma $ is the torsion angle
and $\vec{C}$ is the chiral vector. $\vec{T}_{0}$ and
$\vec{T}_{new}$ are the translational vectors for the undeformed
and the deformed tube.}
\end{figure}

In this part, the torsional strain effect on the Raman modes is
investigated. Here, three tubes with different chiralities are
taken as an example, i.e., armchair (5, 5), zigzag (10, 0) and
chiral (12, 4) tubes. The torsion angle \textit{$\gamma $} is
defined in a similar way to that in Ref. \onlinecite{r30}, i.e.,
it makes any point $\left( {r_c , r_t } \right)$ on the 2D
projection of the tube surface move to $\left( {r_c - tan\left(
\gamma\right)r_t, r_t } \right)$ (see Fig. 6). The following
method is used to find a new periodical structure under torsional
strain: firstly, an origin of the coordinates is selected as the
site $O$, from which, we can define a chiral vector $\vec {C}$ and
the original translational vector $\vec {T}_0 $. Then, we find an
equivalent site $B$ for the site $O$. By projecting the vector
$\overrightarrow {OB} $ to the $\vec {T}_0 $, a new translational
vector $\overrightarrow {O{B}'} $ can be produced as a tube
period, and the angle between $\overrightarrow {OB} $ and
$\overrightarrow {O{B}'} $ is the corresponding torsion angle
$\gamma $. After structure relaxation on both the lattice constant
along the tube axis and the atomic positions, we can finally
obtain a reasonable structure which can be used for further first
principle calculations. And in all of our calculations, the system
keeps a cylinder structure and does not appear any structure
transformations \cite{r57} or defects \cite{r58}.

\begin{figure*}
\includegraphics[width=0.66\columnwidth]{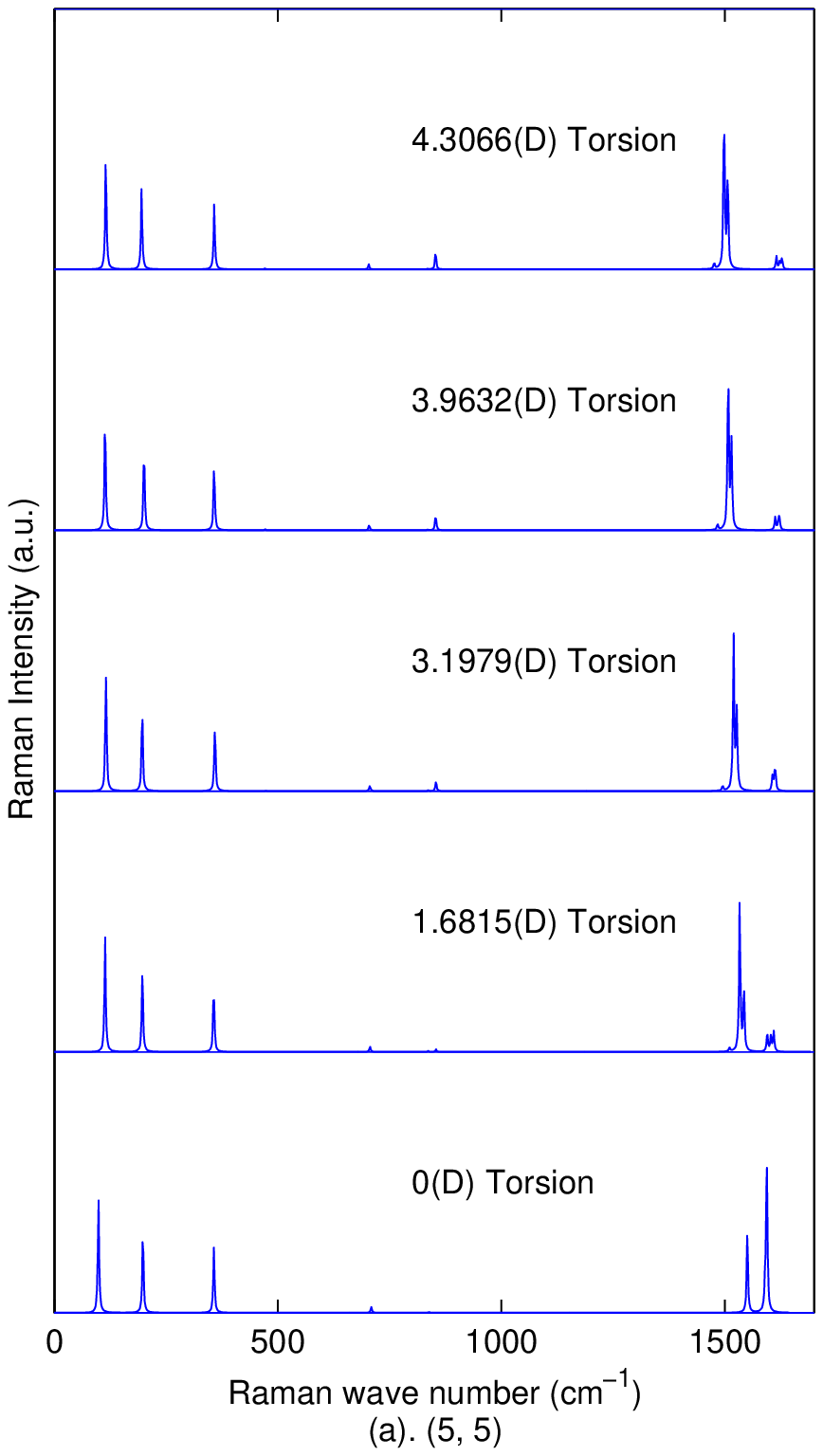}
\includegraphics[width=0.66\columnwidth]{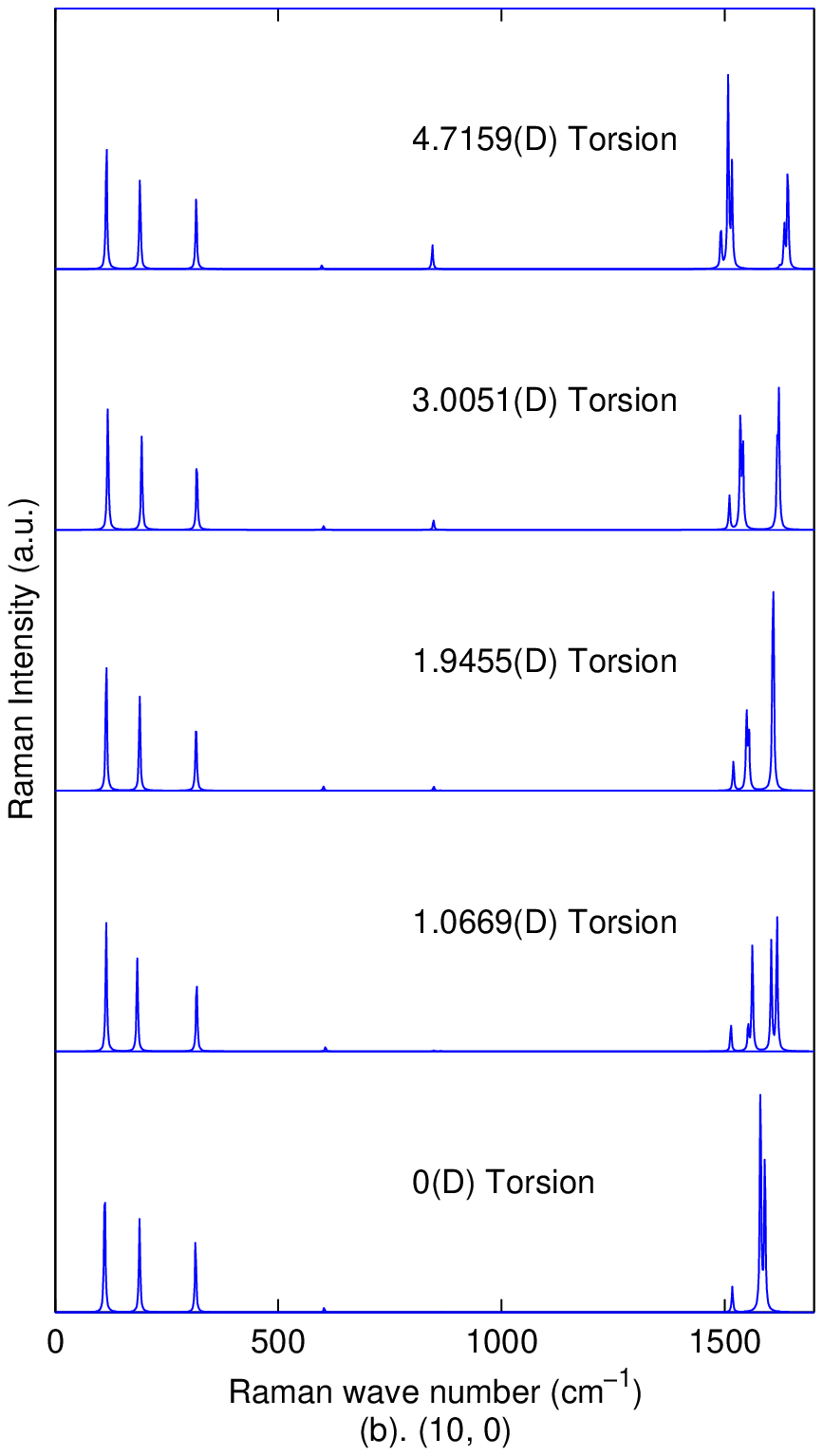}
\includegraphics[width=0.66\columnwidth]{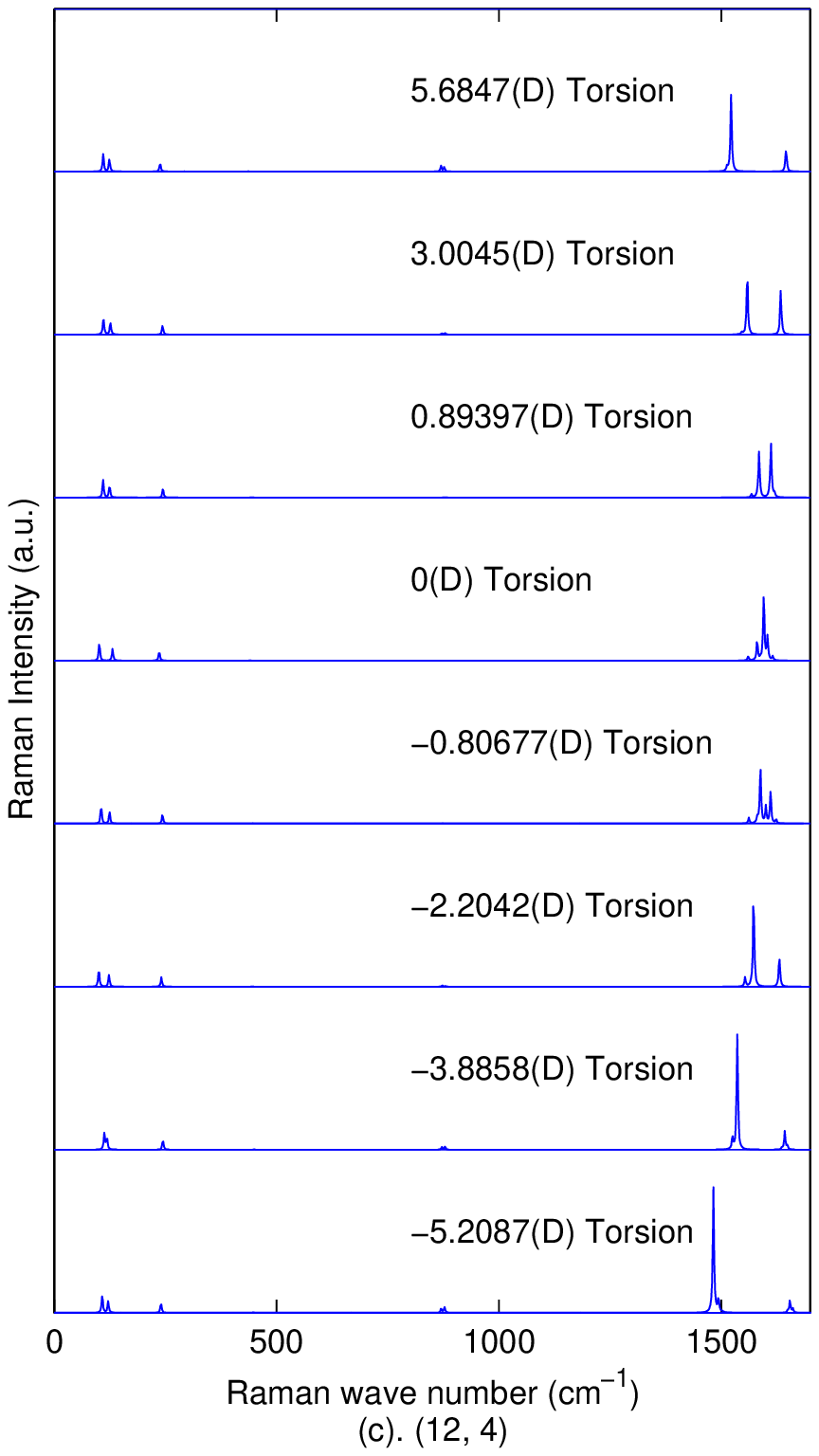}
\caption{\label{fig7}The Raman spectra of different SWCNTs under
torsional strain. (a) (5, 5), (b) (10, 0), (c) (12, 4).}
\end{figure*}

The Raman spectra of different SWCNTs under torsional strain have
been shown in Fig. 7. Because of the symmetry, we only consider
the case of positive $\gamma $ for the armchair and zigzag tubes.
Compared with the situation under the tensile strain, we can find
two similar key characters: a) Not only the frequencies of the
low-frequency Raman active modes, but also their intensities
almost do not change. Some of the high-frequency modes in the
Raman spectra are red-shifted and some ones are blue-shifted. b)
The Raman active vibration number changes under torsional strain.

In fact, the first character can be explained by the same reason
as that under tensile strain. For example, the radii of the
optimized (12, 4) tubes changes only about 0.2{\%} when the
torsional stain is increased up to 6\r{ }. And the two bonds
mirroring each other with $\sigma _v $ symmetry (see Fig. 1 of
Ref. \onlinecite{r53}) will be elongated and shortened under the
torsional strain, respectively, making thus some Raman modes
red-shifted, and some others blue-shifted. The second character
may be not so obvious like the first one. In fact, this character
manifests itself in two different ways, originating from different
mechanisms. The frequencies of Raman active modes for the three
tubes under torsional strains are shown in Fig. 8, from which, it
is seen that in the range from 1400 to 1600 cm$^{-1}$ for all the
tubes, some modes split into two new modes, obviously arising from
symmetry breaking. Since the torsional strains applied in this
work is smaller, the two new modes are very close to each other.

\begin{figure*}
\includegraphics[width=0.66\columnwidth]{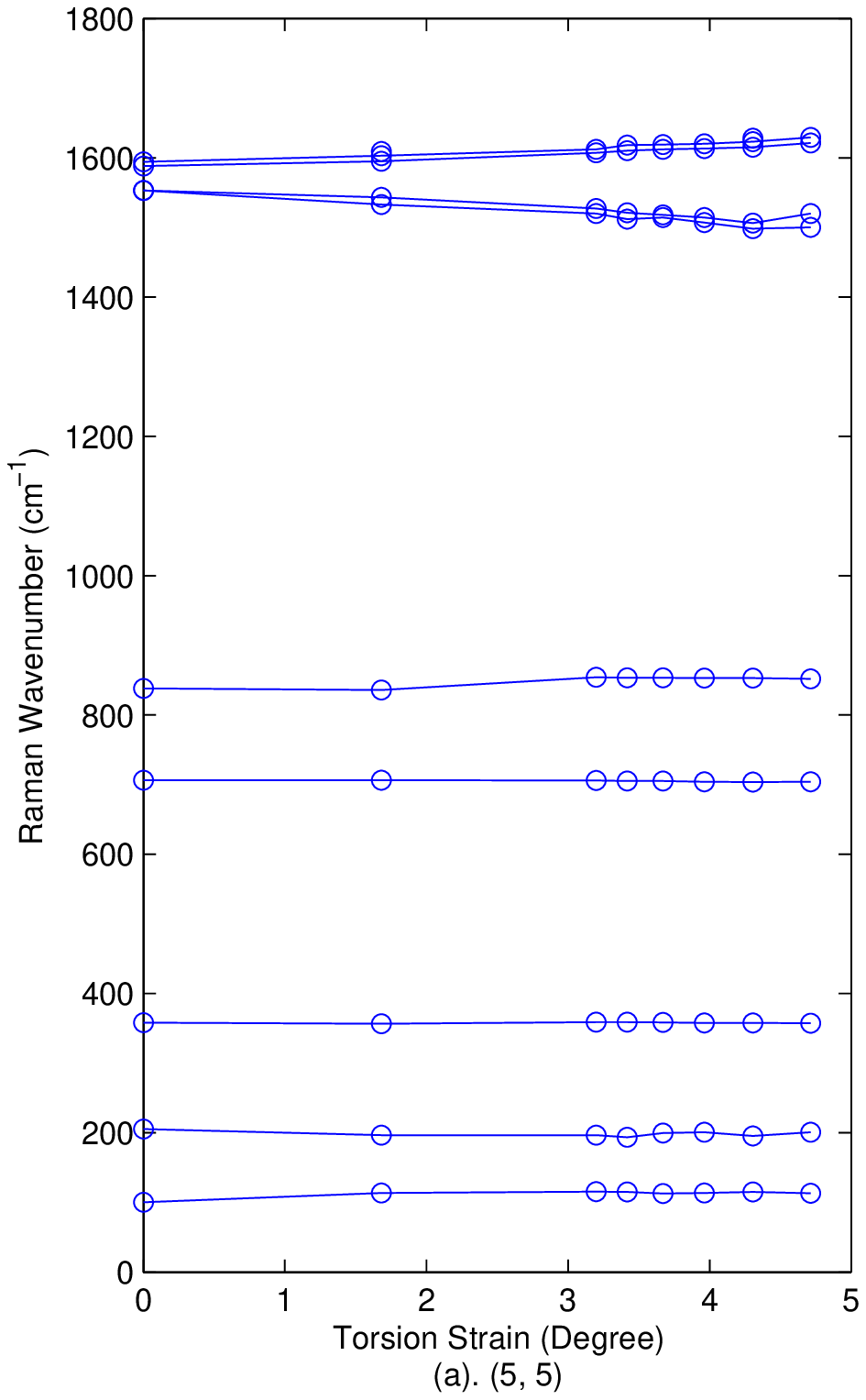}
\includegraphics[width=0.66\columnwidth]{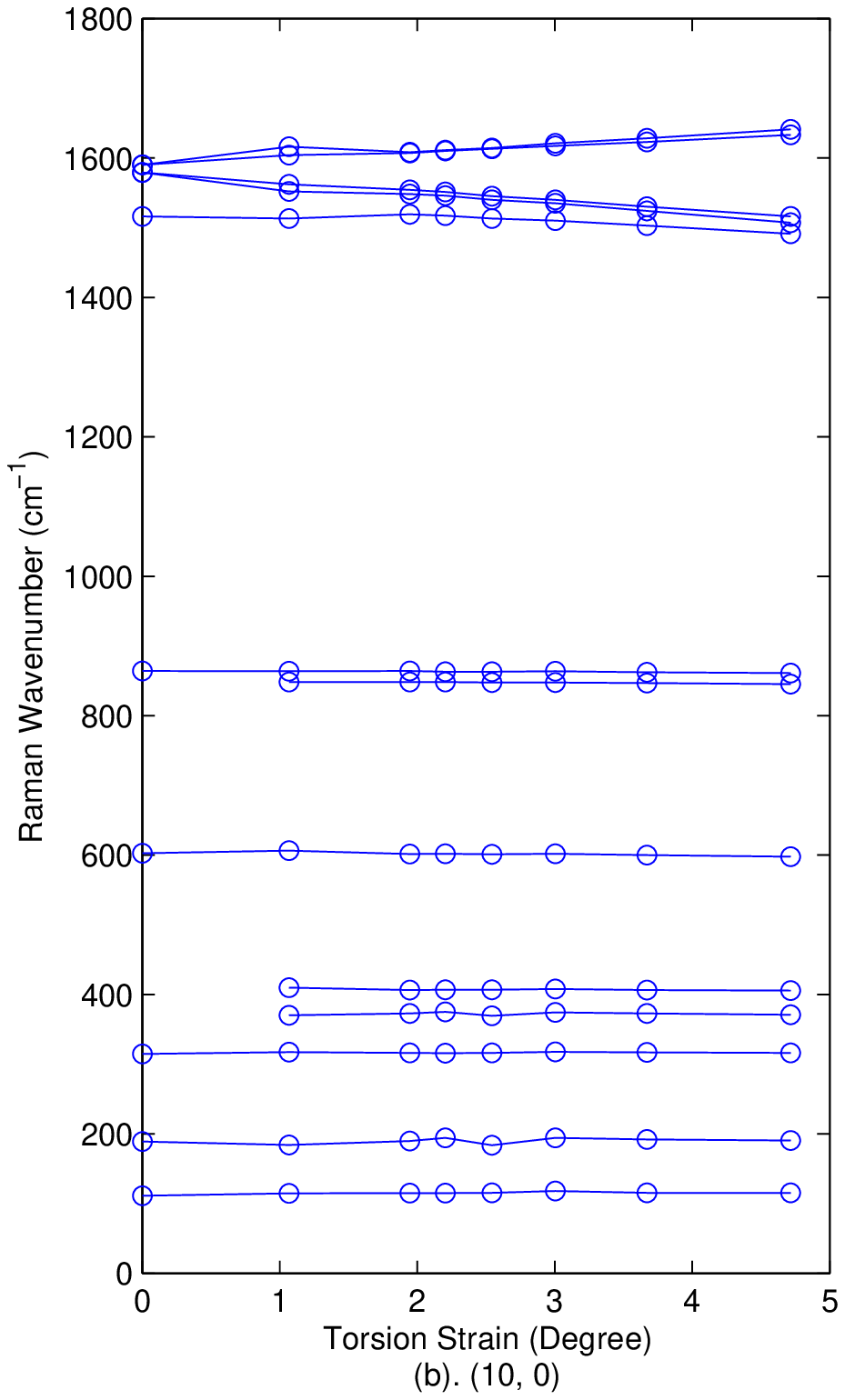}
\includegraphics[width=0.66\columnwidth]{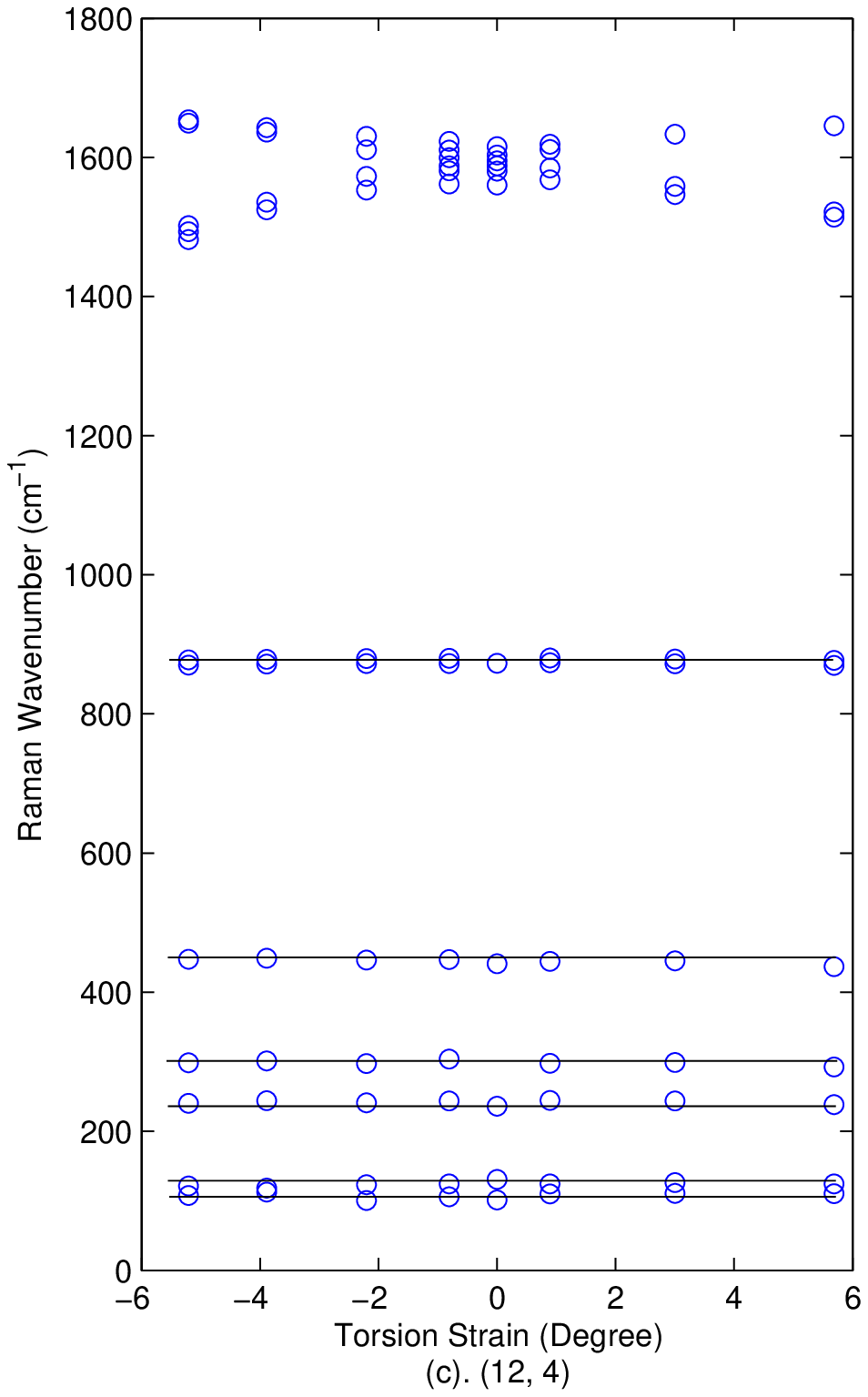}
\caption{\label{fig8}Frequencies of Raman active modes vs
torsional strains for three SWCNTs. (a) (5, 5), (b) (10, 0), (c)
(12, 4).}
\end{figure*}

On the other hand, some new low-frequency modes (about 400 cm$^{ -
1})$ appear in the zigzag tubes, which can not simply explained by
symmetry breaking. In order to find the origin of these new modes,
the separate contributions to the Raman intensities from the three
different terms in Eq. (\ref{eq5}) are investigated. The same
technique had been used in Ref. \onlinecite{r28} to gain a deeper
insight into the significance of the separate contributions due to
different hydrocarbon parameters and the fit parameters for
C$_{60}$. Only the results under torsional strain are given in
Fig. 9 because the three parts vanish in undeformed SWCNTs. From
Fig. 9, one can find that the torsional strain do break the
forbidden rule of these modes\cite{r53} and affect their Raman
intensity dramatically. With torsional strain, the contributions
from the second and the third term in Eq. (\ref{eq5}) enhance
greatly. The second term in Eq. (\ref{eq5}) corresponds to the
contribution from the anisotropic part of the polarizability under
bond stretching, and the third term corresponds to the
contribution from bond rotation. Obviously, the three bonds are
elongated differently under torsional strain, which means the
anisotropic effect will be important. And because torsional strain
makes the bonds rotate, the third term will also enhance.
Altogether, the scattering cross section and thus the total Raman
intensity at a specific frequency will increase under torsional
strain.

\begin{figure}
\includegraphics[width=\columnwidth]{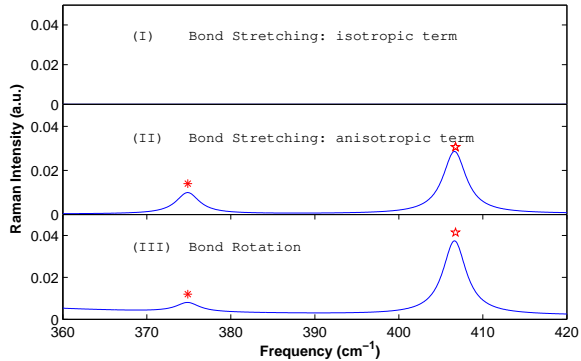}
\caption{\label{fig9}(Color online) Raman peak intensities of
deformed (10, 0) tube with torsional angle of 2.2047\r{},
contributed from the three individual terms in Eq. (\ref{eq5}).
The asterisk and pentagram indicate the positions of new
Raman-active modes. Panel (I), (II) and (III) give the
contributions from the first, the second and the last term in Eq.
(\ref{eq5}), respectively. Notice that the total intensity is not
the sum of those in the three panels due to interference effects
between them.}
\end{figure}

Finally, the GTEs of high-frequency Raman modes of these tubes
under torsional strain are calculated, which are defined in the
present work as $\kappa _t = - \frac{16}{3}\frac{d\ln \omega
}{d\tan ^2\gamma }$. Here, $\omega $ is the frequency of
Raman-active modes, and $\gamma $ is the torsional strain. This
definition is obtained by assuming $\frac{\omega }{\omega _0 } =
\left[ {\frac{r}{r_0 }} \right]^{3\kappa _t }$, which is taken
from Ref. \onlinecite{r54}. The $\kappa _t $ for the 4 highest
Raman-active modes of (5, 5) tube are found to be -26.73, -26.52,
19.85 and 15.65, respectively, and $\kappa _t $ for the 5 highest
Raman-active modes of (10, 0) tube are -13.68, -30.68, -28.37,
17.95 and 19.80, respectively. Because of the complexity of the
high-frequency Raman modes of (12, 4) tube, here only its two most
intense modes at about 1600 cm$^{ - 1}$ are considered, whose
$\kappa _t $ are -32.29 and 14.03. One can find that these values
do not obviously relate to each other, which is different from
those under the tensile strain. So, more investigations should be
done to find the intrinsic mechanism to cause the shifts of the
Raman-active modes under torsional strain.

\section{Conclusions}

In this paper, the nonresonant Raman spectra of chiral and achiral
SWCNTs under uniaxial and torsional strains have been
systematically studied by the \textit{ab initio} calculations and
the empirical bond polarizability model. The cumulant
force-constant method is used to construct the dynamical matrices.
Under tensile strain, it is found that the frequencies and the
intensities of the low-frequency Raman active modes almost do not
change, while their high-frequency part shifts downward linearly
with two different slopes for any kind of SWCNTs. And the Raman
active vibration number does not change under the tensile strain.
Under the torsional strain, the frequencies and the intensities of
the low-frequency Raman active modes still change very little, but
some of the high-frequency modes are red-shifted and some others
blue-shifted. More importantly, new Raman peaks are found in the
nonresonant Raman spectra under torsional strain, which are
explained by a) the symmetry breaking and b) the effect of bond
rotation and the anisotropy of the polarizability induced by bond
stretching.

\begin{acknowledgments}
The authors acknowledge support from the Natural Science
Foundation of China under Grants No. 10474035 and No. A040108. The
authors also thanks support to this work from a Grant for State
Key Program of China through No. 2004CB619004. Gang Wu would like
to thank Dr. Ye Linhui for useful discussions on CFC method. Our
calculations have been done on the Sun Fire V20z computers.

\end{acknowledgments}

\end{document}